\documentclass[twocolumn,english,aps,prb,groupedaddress,amsmath,amssymb,floatfix,reprint]{revtex4-1}

\usepackage{graphicx}
\usepackage{amsmath}
\usepackage{verbatim}
\usepackage{amstext}
\usepackage{amssymb}
\usepackage[version=4]{mhchem}

\usepackage{xcolor}

\def\K{\,\textrm{K}}
\def\eV{\,\textrm{eV}}
\def\meV{\,\textrm{meV}}

\def\diag{\textrm{diag}}
\def\eff{\textrm{eff}}
\def\avg{\textrm{avg}}

\newcommand{\angstrom}{\,\mbox{\normalfont\AA}}

\definecolor{lb}{rgb}{0.000, 0.500, 1.000}

\begin{document}

\title{First-Principles study of an $S=1$ quasi-1D quantum molecular magnetic material}
\author{Maher Yazback}
\author{Jie-Xiang Yu}
\author{Shuanglong Liu}
\author{Long Zhang}
\author{Neil S. Sullivan}
\author{Hai-Ping Cheng}
\email{hping@ufl.edu}

\affiliation{Department of Physics, University of Florida, Gainesville, Florida 32611, USA}
\date{\today}

\begin{abstract}
% Maher's version of abstract. 
We use density functional theory to study the structural, magnetic and electronic structure of the organo-metallic quantum magnet $\mathrm{NiCl_2}\textrm{-}\mathrm{4SC(NH_2)_2}$ (DTN). 
Recent work has demonstrated the quasi-1D nature of the molecular crystal and studied its quantum phase transitions at low temperatures. 
The system includes a magneto-electric (ME) coupling and, when doped with Br, the presence of an exotic Bose-glass state. 
Using the generalized gradient approximation (GGA) with inclusion of a van der Waals term to account for weak inter-molecular forces and by introducing a Hubbard $U$ term to the total energy, we systematically show that our calculations reproduce the magnetic anisotropy, the inter-molecular exchange coupling strength and the magneto-electric effect in DTN, which have been  observed in previous experiments.
Further analysis of the electronic structure gives insight into the underlying magnetic interactions, including what mechanisms may be causing the ME effect. 
Using this computationally efficient model, we predict what effect applying an electric field might have on the magnetic properties of this quantum magnet.
\end{abstract}
\maketitle

\section{Introduction}
Interest in materials that exhibit sizable magneto-electric (ME) effects has grown considerably within the past decade. New classes of materials exhibiting the phenomena hint at applications where it can be harnessed, and materials can be designed for use in potential low-power spintronic devices\cite{spintronic} and future computing applications. The ME effect is characterized by a coupling between the magnetization and electric polarization of a material, 
that is, the application of a magnetic field induces a change in the electric polarization of a material and similarly, the application of an electric field causes a change in the magnetization. Investigations into the ME effect have largely focused on its presence within transition-metal oxides\cite{multiferroic}.
Recent studies have shown that this coupling between the electric and magnetic properties may also be present within the class of materials known as organo-metallic molecular crystals\cite{multiferroic}.

The fundamental building block of these solids is a molecule where individual molecular units are held together by relatively weak inter-molecular interactions. 
At the center of each molecule is a magnetic metal ion, and the interactions between neighboring magnetic moments dictate the magnetic properties of the crystal. 
Given the weak interactions between molecules, molecular crystals are often easily strained under external magnetic and electric forces\cite{magnetostriction}. 
This soft lattice structure may provide a degree of freedom through which the ME effect may be tuned. 
The potential to synthesize organic ligands may also allow future flexibility in designing and tuning properties of such materials\cite{sato2016dynamic}.

\begin{figure}
\includegraphics[width=1\columnwidth]{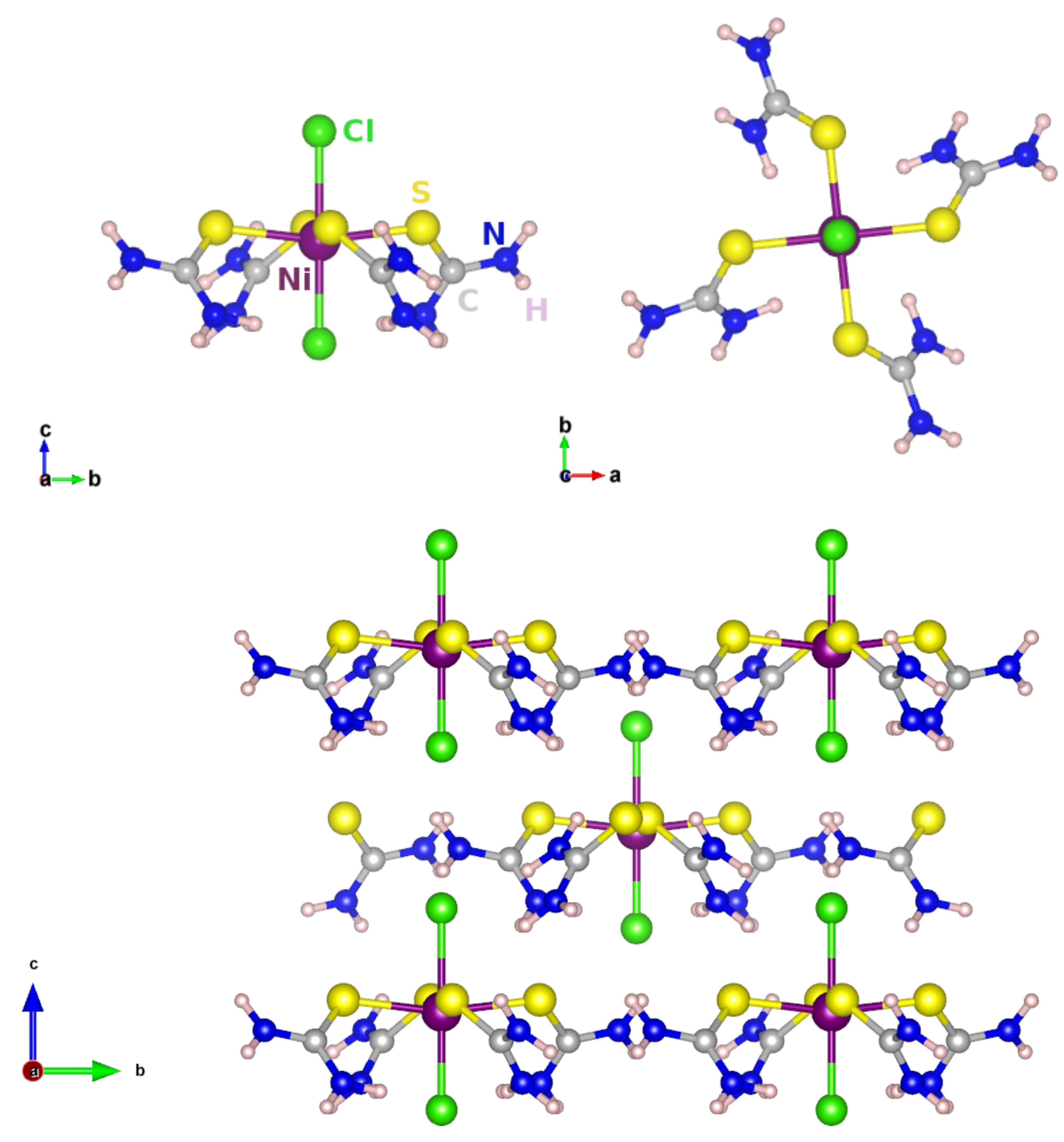}
\label{fig:structure}
\caption{
(Top) Two views of the the molecular unit making up the DTN molecular crystal. Four thiourea ligands surround the magnetic Ni ion. (Bottom) The DTN unit cell.
%\JX{Color code: pink Ni, yellow S, green Cl, blue N, grey C, white H.}
} 
\end{figure}

The subject of this work is the organic quantum magnetic system dichloro-tetrakis-thiourea nickel,  $\mathrm{NiCl_2}\textrm{-}\mathrm{4SC(NH_2)_2}$ (DTN),\cite{structure} which has been studied in experiments and by quantum Monte-Carlo simulations based on model Hamiltonians \cite{originOfCoupling,multiferroic,nature,magnetostriction,nmrDTN,DTNX, DTNreview}. The molecular unit of this system (Figure~\ref{fig:structure}) has a central magnetic Ni cation surrounded by four electrically polar ligands, $\mathrm{C(NH_{2})_{2}S}$ (thiourea). 
DTN and its doped derivatives have been studied extensively due to the rich physics present at low temperatures. The phase diagram\cite{multiferroic,nature,magnetostriction} of pure DTN shows that, at temperatures below $ 1.2 \K $ and below a critical magnetic field, $H_{c1}$, it is a quantum paramagnet. As the magnetic field, applied perpendicular to DTN's $ab$-plane, reaches the first critical field, it experiences a quantum phase transition into an $XY$-antiferromagnetic ($XY$-AFM) state where all spins now lie within the $ab$-plane. 
As the field is increased further the spins begin to cant, with a corresponding increase in magnetization. 
When a second critical field, $H_{c2}$, is reached, the magnetization saturates and the material enters a spin-polarized state where all spins are pointing along the $c$-axis and parallel to the applied magnetic field. 
Within this $XY$-AFM region DTN exhibits the ME effect where, along with an increase in magnetization, there is a correlated increase in the electric polarization. 
In this paper we investigate the structural, electronic and magnetic properties of DTN from first principles. 
Using appropriate levels of approximation within density functional theory \cite{hohenbergKohn, kohnSham} (DFT) we are able to balance accurate structural predictions with computational efficiency to gain insight into DTN's electronic and magnetic structure, the possible mechanisms responsible for its ME effect, and investigate properties of the solid not yet studied in the literature. 

The rest of the paper is organized as follows. Section II A presents an overview of the computational details involved in the calculations. 
%as implemented within the software package Vienna Ab initio Simulation Package (VASP) \HPC{reference}. 
Section II B discusses the theory behind modelling dispersion interactions and the approximations we made in attempting to obtain accurate structural properties. In section II C we discuss the addition of $U_{\eff}$ and the justification for our choice. In Section III we discuss our results relating to the systematic improvements made to our model, comparing to experiments along the way. Finally we go on to predict experimentally unverified properties of the material and summarize our conclusions in Section IV.
% {\color{red}HPC: It is not a good idea to put others figures in your introduction. % % You can refer to them while you talk about their findings}
% DTN molecule

% DTN phase diagram 
% Is this phase diagram your work? --JX
%\begin{figure}[h]
%\centering
%\includegraphics[width=0.42\textwidth]{bose-glass-phase-diagram.png}
%\caption{\label{Figure 2.} 
%The phase diagram for pure and doped DTN. Doping with 8\% Br introduces the exotic Bose-glass phase.}
%\end{figure}

% Zeeman splitting figure
%\begin{figure}[h]
%\centering
%\includegraphics[width=0.42\textwidth]{energy-splitting.png}
%\caption{\label{Figure 3.}
%a) Energy-level splitting of $S_z=0, \pm1$ of pure DTN under the application of a magnetic field. b) Spin canting in the direction of the applied field.   
%\cite{}}
%\end{figure}

% Magneto-electrcic coupling 
%\begin{figure}[h]
%\centering
%\includegraphics[width=0.42\textwidth]{magnetoelectric-effect.png}
%\caption{\label{Figure 4.} 
%Experimental measurements of the magneto-electric coupling. Inset shows strain induced %by magnetostriction, suspected to be the cause of the coupling. 
%\cite{}}
%\end{figure}

\section{Methodology}\label{sec2}

\subsection{Overview}
% overview

\begin{figure}
\includegraphics[width=1\columnwidth]{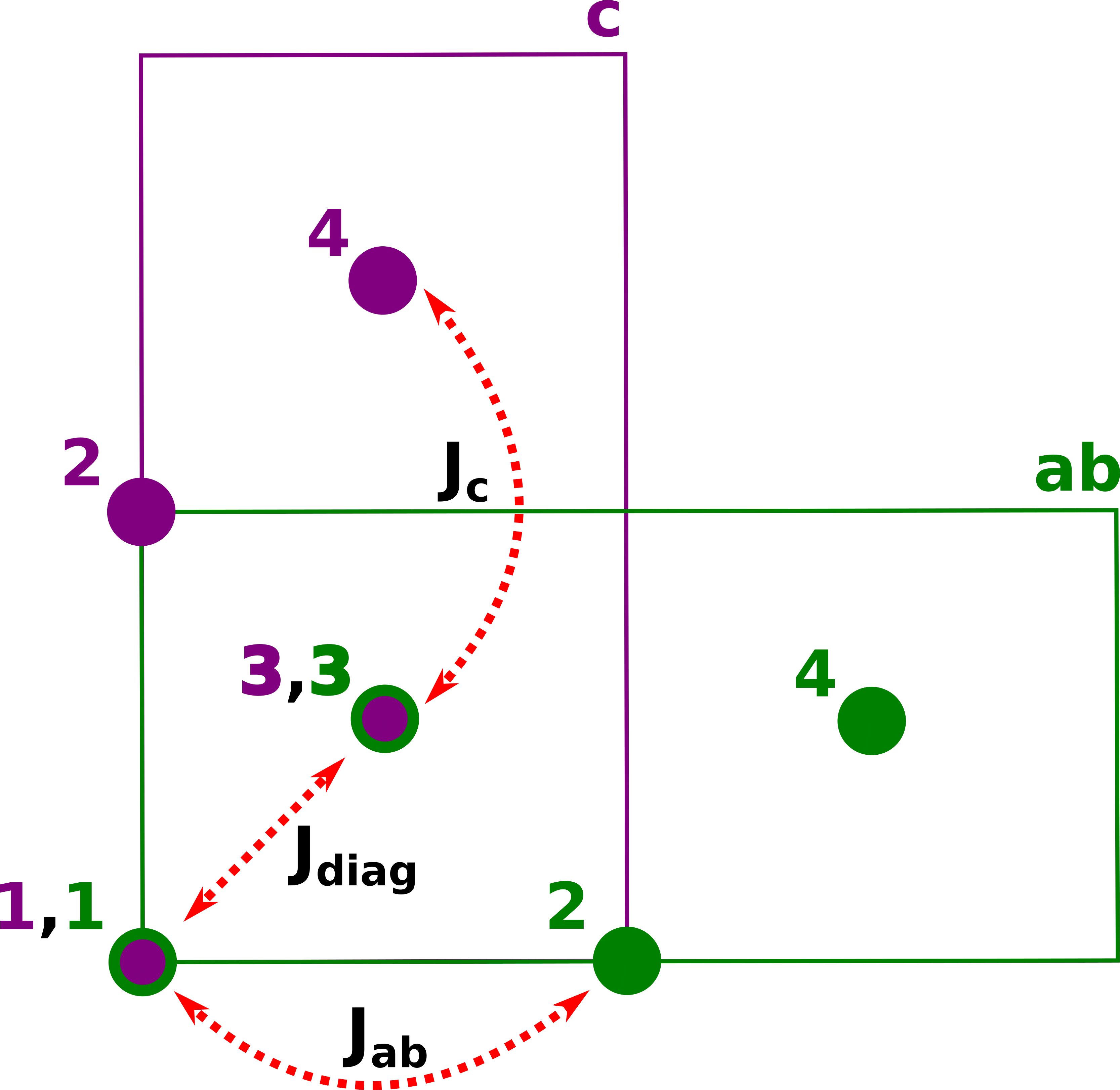}
\caption{
The $2\times1\times1$ ($ab$, green) and $1\times1\times2$ ($c$, purple) supercells used to calculate respective exchange coupling $J$ values.
The nodes correspond to Ni ions and red arrows indicate the exchange interactions.}
\label{fig:supercell} 
\end{figure}

We performed first-principles calculations using the Vienna \textit{ab initio} Simulation Package (VASP)\cite{vasp1,vasp2,vasp3} within the density functional theory (DFT) framework.

The projector augmented wave pseudopotentials\cite{PAW_1994,PAW_1999} and generalized gradient approximations (GGA) of Perdew, Burke, and Ernzerhof \cite{PBE} were used for the exchange-correlation energy.
% +U
To give an appropriate description of the on-site Coulomb interaction between Ni($3d$) electrons, we apply the plus $U$ method \cite{DFTU} (GGA$+U$) with $U_{\eff}=U-J=6.0$~\eV~on Ni when investigating the exchange couplings and the magnetic anisotropy of DTN.
% vdW, just mention DFT-D3.
The D3\cite{dftd3} dispersion correction combined with GGA (GGA-D3) and/or GGA+$U$ (GGA-D3$+U$) was used to take into account Van der Waals (vdW) interactions. The damping parameter for D3 was set to $S_R=1.5$.
% ENCUT
Wavefunctions were expanded in plane waves and an energy cutoff of $520 \eV$ was used for all calculations. 

% K Mesh
The Brillouin zone of the DTN unit cell containing two $\mathrm{Ni}^{2+}$ ions was sampled on a $6\times6\times6$ Monkhorst-Pack\cite{monkhorstpack} mesh for total energy and structure optimizing calculations with an applied Gaussian smearing of $ 0.01 \eV$. 
% PDOS
A $10\times10\times10$ $\Gamma$-centered $K$-mesh with the tetrahedron smearing method was used for the density-of-state (DOS) and orbital-resolved projected DOS (PDOS) calculation.
% relax
The structure of DTN was relaxed for both the lattice constant and the atomic positions until atomic forces on each atom were converged to within $ 0.01 \eV/\angstrom $.

% %
% \JX{DELETE:The Projector-augmented wave potentials\cite{vasp4,vasp5} were used and convergence threshold was set to $10^{-8}$ eV.} 
% %
% \JX{DELETE:The DTN structure was relaxed using the PBE functional\cite{PBE} with the GGA-D3 dispersion correction and damping set to $S_R=1.5$, which . \MY{The GGA-D3 correction contributes only a structure-dependent correction to the total energy but was applied to all calculations, whether structural relaxation was needed or not}. Forces were converged to within 0.01 eV/atom. Comparisons between different functionals on the structure are discussed in Section III-A. }

% J
Calculations of the exchange coupling $J$ between magnetic Ni ions were performed by creating a supercell composed of two DTN unit cells with four Ni atoms, as shown in Figure~\ref{fig:supercell}.
% J_{c} and J_{ab}
In order to compare to experiment, we focused on the $J_{c}$ coupling along the $c$-axis of the unit cell and $J_{ab}$ coupling between Ni ions within the $ab$-plane, or along the $a/b$ axis. 
% J_{diag}
In addition, the $J_\diag$ coupling, which connects the corner Ni with the body-centered Ni ion was also considered.
% J for supercell
Two supercells, one of $1\times1\times2$ extending along $c$-axis and the other of $2\times1\times1$ extending along $a/b$-axis, were created to isolate inequivalent Ni ions. 
% K mesh for supercell
The corresponding $K$-mesh was therefore $6\times6\times2$ or $2\times6\times6$, respectively.

% \JX{DELETE:
% \MY{All exchange coupling calculations were performed using co-linear spins and neglecting spin-orbit coupling.}
% }
% %
% The total energy differences of different colinear spin configurations 
% We fit $J$'s to the Heisenberg model and compare results to experiment for GGA and GGA+$U$. 

%MLWF

In order to gain insight into the magnetic interactions in DTN, we transformed the plane-wave-based Bloch states from VASP into Wannier function- (WF-) based local states by using the maximally-localized Wannier functions method \cite{mlwf} implemented in the Wannier90 package\cite{wannier90}. 
% downfolding
By choosing an initial guess of projections from the outer energy window, an effective tight-binding Hamiltonian $\mathcal{H}$ was built in a downfolded Hilbert subspace where the eigenvalues in the inner energy windows are exactly the same as those of DFT results\cite{MLWF_DIS}. 
% result
Microscopic insights into the relevant orbitals and superexchange paths in DTN are based on the hopping integrals and on-site energies of the WF-based Hamiltonian, $\mathcal{H}$. 

% calculated maximally-localized Wannier functions (MLWF) using the wannier90 software package\cite{wannier90}. Wannier functions (WF) are an analog of molecular orbitals for solid state materials. Unlike Bloch states which are de-localized and defined by a band index $n$ and crystal momentum $\textbf{k}$, Wannier functions are spatially localized and centered on a lattice $\textbf{R}$ \cite{mlwf}. Given the large number of Bloch states in Hilbert space resulting from a groundstate DFT calculation, this transformation to a WF basis can be performed by choosing a subset of Bloch states forming bands near the Fermi level and Fourier transforming them into the new, smaller basis set. The software follows the method of Marzari and Vanderbilt\cite{mlwf} for minimizing the spread of MLWF. Our calculations used eight bands near the Fermi level. 

%SOC anisotropy
Magnetic anisotropy is a consequence of spin-orbit interactions. 
We therefore performed non-colinear magnetic DFT calculations including spin-orbit interactions.
% constrained configuration
We obtained the total energies for various configurations with each having a set of constrained directions of the local magnetic moments for the Ni atoms.
% MAE
The resulting total energy difference between configurations with moments perpendicular to one another gave the magnetic anisotropic energy.
%easy axis
The direction with the lowest total energy is usually labeled as the easy axis.

\subsection{Test of Van der Waals correction}

% Structural relaxation table
\begin{table}[h]
\caption{
Lattice constants of DTN with optimized structures obtained using different exchange-correlation functionals.
% \JX{
% COMMENT: it is a chaos here. Why LDA and PBEsol and HSE appear here? Did we introduce them even one sentence in Method section? My suggestion is we just leave GGA, GGA-D3, optB88 and DF2, then we can say we investigate vdW interaction, then the discussion of vdW is meaningful.
%}
}
\begin{ruledtabular}
\begin{tabular}{ c  c  c  }
 XC Functional & $a$ (\AA) & $c$ (\AA) \\ 
 \hline
% LDA          & 9.29  & 8.44 \\
 GGA          & 9.70  & 9.35\\
 GGA-D3 ($S_{R}=1.2$) & 9.55  & 8.78 \\
 GGA-D3 ($S_{R}=1.5$)      & 9.58  & 8.92\\
 DF2          & 9.77  & 9.02\\
 optB88       & 9.54  & 8.63\\
% PBEsol       & 9.41  & 9.07\\
% HSE06        & 9.67  & 9.16\\
 \textbf{Experiment}\cite{structure}  & 9.56  & 9.08\\
 \end{tabular}
\end{ruledtabular}
\label{table:lattice-constants}
\end{table}

Molecular crystals pose a challenge to theorist when trying to predict their structural properties. 
This has been especially true for GGA exchange-correlation functionals in DFT calculations, which struggle to account for long-range correlations like the Van der Waals (vdW) force \cite{vdW} and hydrogen bonding.
New functionals, optB88 \cite{optB88} and DF2 \cite{DF2}, have been introduced recently to model these dispersive interactions accurately. 
In order to balance the computational cost of calculations while still accurately predicting molecular structures, we also consider the D3\cite{dftd3} dispersion correction, which is frequently cited for successfully predicting molecular crystal structures. The D3 dispersion adjusts the inter-nuclear energy to account for the long-range asymptotic behaviour with little effect at short range, giving an augmented DFT total energy.

We chose the vdW correction for our study after performing a benchmark test for the lattice constants of DTN.
The DTN unit cell with two Ni atoms was used. 
The co-linear magnetic moments on the two magnetic Ni atoms in the unit cell are aligned anti-parallel. 
Structures were optimized for both lattice and atomic positions.
The lattice constants for the DTN molecular crystal were investigated with various functionals and were compared with experimental data. 
The resulting tetragonal lattice constants, $a$ and $c$, are listed in Table~\ref{table:lattice-constants}. 
% GGA-PBE
Without any vdW correction, both the lattice constants, $a$ and $c$, are severely overestimated while using the GGA functional,  
the $c$ lattice constant is especially, being $ 0.27 \angstrom $ larger than the experimental value, indicating the failure to consider long-range vdW interactions with GGA.
% optB88
Among the three vdW corrections GGA-D3, optB88 and DF2, optB88 underestimates lattice constants, especially $c$ with a $ 0.45 \angstrom $ difference compared to the experimental values. 
% DF2
DF2 has small deviations, $ 0.21 \angstrom $ for $a$ and $ 0.06 \angstrom $ for $c$, respectively, compared to the experimental values. 
% GGA-D3
GGA-D3 with the default value of damping parameter $S_R=1.217$ overestimates the vdW interaction, so that it has a relatively large deviation of $ 0.30 \angstrom $ for $c$, compared to the experimental value.
% GGA-D3 with SR=1.5
GGA-D3 with a damping parameter $S_R=1.5$ gives the best agreement with the experimental data. 
The deviation is only $ 0.02 \angstrom $ for $a$ and $ 0.16 \angstrom $ for $c$. 
% conclusion
Considering that the D3 dispersion correction has less computational cost and performs better than optB88 and DF2, we employed GGA-D3 and GGA-D3$+U$, with the adjusted damping parameter, for our future calculations on DTN. %For comparison, GGA-D3 plus $U-J$ is also tested for one set of $U$ and $J$ values \cite{RN3269}.
The lattice constants of DTN were set to $a=9.58 \angstrom $ and $c=8.92 \angstrom $, the GGA-D3 result, for the rest of our calculations.

\subsection{Estimation of $U_{\eff}$}

Constrained Random Phase Approximation (cRPA) calculations \cite{Aryasetiawan_2004a, Aryasetiawan_2006a} for the Ni site were carried out using the full-potential linearized augmented plane wave method implemented in the Exciting-Plus code \cite{Kozhevnikov_2010a} to have a reasonable estimate of the on-site Coulomb interaction of the Ni($3d$)-orbitals.

The basic idea of cRPA is to calculate a partial RPA particle-hole polarization with the constraint of a physically motivated correlation window, for example the $d$-like bands around the Fermi level.
The on-site screened Coulomb interaction strength then can be determined from the partial RPA particle-hole polarization and the bare Coulomb interaction. 
The RPA polarization is obtained from Kohn-Sham susceptibility, which is completely based on the DFT ground state. 
We benchmarked it against other implementations using late transition metal monoxides and got consistent results\cite{Zhang_2019a}.

The molecular unit of DTN, as shown in Figure~\ref{fig:structure}, that was used has the Ni sitting in the center of the octahedron formed by four S atoms and two Cl atoms.
A unit cell with bulk lattice constants but containing a single DTN molecule with 35 atoms in total was taken into account.
This is enough for the calculation of the Kohn-Sham susceptibility and partial RPA polarization because the molecule is well separated from its nearest neighbors in the diagonal directions, and the feature of the strong interaction along the $c$-axis is preserved.
The resulting diagonal elements of the $U$ matrix, the on-site $U$, of $d$ orbitals is $\thicksim 4.9 \eV $ and $J \thicksim 0.5 \eV $.
The cRPA calculation scheme is based on a paramagnetic ground state and magnetic properties of the material are not taken into account, 
so that the Pauli exclusion principle is not included in the calculation.
This ignorance causes underestimation of the $U$ values\cite{Zhang_2019a}. 
Thus, $U=4.9~\eV$ is considered as a lower bound for the value of $U_{\eff}$ for our GGA-D3$+U$ calculations.

% U dependent J

To further understand the influence of $U_{\eff}$ on the results of our GGA-D3$+U$ calculations, we calculated $J_{c}$, couplings along the $c$-axis, for various $U_{\eff}$ values. 
We used the structures obtained with the GGA-D3 functional for this benchmark test.
%result
The results are shown in Table~\ref{table:average}. %\HPC{where is the figure?}
The value of $J_{c}$ for $U_{\eff}=4.9$ and $J=0.52 \eV$ (calculated by cRPA) is $-0.31 \meV$ and for $U_{\eff}=4.9$ and $J=0.0 \eV$ is $-0.28 \meV$. The experimental value \cite{nature} is $-0.19 \meV$ ($-2.2 \K $) where the negative value corresponds to an antiferromagnetic coupling. This is a significant improvement over the GGA-D3 value of $-1.05 \meV$. Since $U-J$ only adds a small correction, for the rest of the test we set $J=0.0 \eV$ for simplicity: Then the values of $J_{c}$ for $U_{\eff}=6.0$ and $7.0 \eV$ are $-0.21$ and $-0.16\meV$, respectively.
Considering the $U_{\eff}$ values used in the previous experimentally validated bulk NiO calculations\cite{DFTU,cai2009study} and the results mentioned previously, we finally chose $U_{\eff}= 6.0 \eV$ for our GGA-D3$+U$ calculations.

% U dependent anisotropy

The $U_{\eff}$-dependent magnetic anisotropy energies were also obtained with various $U_{\eff}$ values.
The energy difference between spins along the $c$-axis and spins along $a$-axis is 0.78--$0.79$ meV (9.05--$ 9.16 \K$) per unit cell for GGA-D3+$U$ with $U_{\eff}=4.9$--$7.0 \eV$, indicating it is insensitive to $U_{\eff}$. Here we set $J=0 \eV$. 
For completeness we calculated the magnetic anisotropy using $U=4.9$ eV and $J=0.52$ eV giving an anisotropy value within the $ 0.78 - 0.79 \meV $ window, which will be discussed again in Section III C.

\section{Results and Discussion}\label{sec3}

\subsection{Structural properties}

The Ni atoms in DTN form a body-centered tetragonal lattice so that there are two Ni atoms, one at the corner and the other at the body-center, in each tetragonal unit cell of DTN.
Each Ni, with a +2 valence state, is surrounded by four S atoms in the $ab$-plane and two Cl atoms along the $c$-direction, leaving them in an octahedral crystal field.
In each $\mathrm{Ni S_4 Cl_2}$ octahedron, Cl atoms are located exactly on the apical sites along the $c$-axis, while S atoms in the polar thiourea ligands, $\mathrm{C (NH_2)_2 S}$, deviate a little from the $ab$-plane containing the central Ni.

A comparison of bond lengths and bond angles between the DFT relaxed structure using GGA-D3 and experiment\cite{structure} is given in Table~\ref{table:bondlengthsandangles}. 
Bond lengths are all within $ 0.02 \angstrom $ and bond angles are within $ 2.5^{\circ}$ of the experimental values. 
This further confirms the consistency between the GGA-D3 approach and experiment.

% \avg J coupling
\begin{table}
\caption{
Average exchange coupling $J^\avg$ in units of meV along the $c$-axis and in the $ab$-plane 
using GGA and GGA+$U$, compared to the experimental results, 
% \HPC{Put all negative signs in the tables between two \$ signs so it looks like $-1$, not -1}
% \JX{done}
}
%\JX{COMMENT: transfer all data into meV}
%}\MY{Done}
\begin{ruledtabular}
\begin{tabular}{c r r r}
  XC & $J_c^{\avg}$ & $J_{ab}^{\avg}$ & $J_\diag$\\
\hline
 GGA-D3            & $-1.05$  & $-0.149$ & $-0.033$ \\
% GGA+D3 (U=4.9 J=0.52)  & -0.31 & -0.027 & -0.016 \\
GGA-D3$+U$ (4.9)   & $-0.28$ & $-0.023$ & $-0.010$ \\
GGA-D3$+U$ (6.0)   & $-0.21$ & $-0.018$ & $-0.008$ \\
GGA-D3$+U$ (7.0)   & $-0.16$ & $-0.015$ & $-0.006$ \\
Experiment         & $-0.19$ & $-0.016$ & N/A\\
% GGA-D3            & -12.2  & -1.73 & -0.38 \\
% GGA-D3$+U$ (4.9)          & -3.19 & -0.27 & -0.12 \\
% GGA-D3$+U$ (6)          & -2.44 & -0.21 & -0.09 \\
% GGA+U (7)          & -1.91 & -0.17 & -0.07 \\
% Experiment     & -2.2 & -0.18 & N/A\\
\end{tabular}
\end{ruledtabular}
\label{table:average}
\end{table}

% Bond length table
\begin{table}[h]
\caption{
Bond lengths and bond angles of structurally relaxed DTN using GGA-D3 compared to experiment.}
\begin{ruledtabular}
\begin{tabular}{ c r r }
 Bond Length (\AA) & Experiment & Calculated  \\ 
 \hline
 Ni-Cl(1)        & 2.40  & 2.41   \\
 Ni-Cl(2)        & 2.52  & 2.54 \\
 Ni-S & 2.46  & 2.44  \\
 S-C         & 1.73 & 1.72  \\
 C-NH$_2$(1)     & 1.34  & 1.34  \\
 C-NH$_2$(2)     & 1.32  & 1.33  \\
    \hline
 Bond Angle (degrees) &  &  \\ 
 \hline
 S-Ni-Cl(1)        & 96.7  & 96.8   \\
 S-Ni-Cl(2)         & 83.3  & 83.2 \\
 Ni-S-C & 113.9  & 114.2  \\
 S-C-NH$_2$ (1)         & 116.9 & 119.4  \\
 S-C-NH$_2$(2)      & 122.3  & 121.3  \\
 NH$_2$(1)-C-NH$_2$(2)     & 120.8  & 119.2  \\
\end{tabular}
\end{ruledtabular}
\label{table:bondlengthsandangles}
\end{table}

\subsection{Exchange coupling}

% A large part of our efforts went into getting a better microscopic understanding of the magnetic interactions at play in DTN.

Each magnetic Ni ion in molecular crystal DTN has spin $S=1$.
The tetragonal DTN bulk has two nonequivalent Ni-Ni spin couplings, one along the $c$-axis and another in the $ab$-plane.
The corresponding spin Hamiltonian in terms of Ni magnetic spins is 
\begin{eqnarray}
\mathcal{H}_{s} &=& -J_c\sum_{\left<i,j\right>}{S_{i}\cdot S_{j}} - J_{ab}\sum_{\left<l,m\right>}{S_{l} \cdot S_{m}} \nonumber \\
    &&{}+ D\sum_{i}{(S_i^z)^2} - g \mu_B H \sum_{i}{S_i^z}
\label{eq:hs}
\end{eqnarray}
where $\left<i,j\right>$ and $\left<l,m\right>$ are the out-of-plane and in-plane neighbors, respectively, $D$ is the coefficient of on-site uniaxial magnetic anisotropy, also known as the zero-field splitting, and $g \mu_B H$ is the interaction with an external magnetic field.

Experimentally it has been found that the exchange coupling $J_{c}$ between Ni ions along the $c$-axis is an order of magnitude larger than the coupling $J_{ab}$ within the $ab$-plane \cite{nature}. 
It is thus suggested that spins in DTN behave as quasi-1D spin chains.
Therefore, we investigated the exchange coupling using the molecular crystal structures obtained from GGA-D3.
As described in the methods section, the calculations for the exchange coupling $J$ values involved creating two supercells, shown in Figure~\ref{fig:supercell}, extended along $ab$ (green) and $c$ directions (purple), to isolate the couplings of interest. We computed DFT total energies for various inequivalent spin configurations and fit the exchange coupling $J$ parameters to the Heisenberg model.
% \HPC{Marher needs to work on this}(we need add a reference for how to calculate $J$ and using one sentence to introduce it. Then we can say how the numerical error comes from and how to get average values and deviation.)
Table \ref{table:average} gives the calculated results for the exchange coupling $J$ for GGA-D3 and GGA-D3$+U$ (with $U_{\eff}=6.0 \eV $) compared to the experimental data.
% overview
Both experiments and theory show that $J_{c}$ is negative, an antiferromagnetic coupling, with strength an order of magnitude larger than $J_{ab}$.

% GGA-D3 Jc
Compared with the experimental value of $-0.190\meV$ ($-2.2 \K$), $J_{c}$ with GGA-D3 is severely overestimated to be $-1.05\meV$ ($-12.2 \K$). 
% GGA-D3 Jab
The calculated $J_{ab}$ is $-0.149\meV$ ($-1.73 \K$) using GGA-D3, which is also an order of magnitude larger than  the experimental value $-0.016\meV$ ($-0.18 \K$). 
% GGA-D3+U
On the other hand, the GGA-D3$+U$ with the adjusted $U_{\eff}$ value gives results consistent with the experimental values for both $J_{ab}$ and $J_{c}$.
% Jdiag
Although the two Ni ions are nearest neighbors, values of $J_\diag$, the coupling between a corner Ni and a body-centered Ni, are even weaker than $J_{ab}$ for both functionals.
% conclusion
Because of the negligible values of both $J_{ab}$ and $J_\diag$, the strong $J_{c}$ coupling makes DTN a quasi-1D antiferromagnetic chain along the $c$-axis.

\begin{figure}
\includegraphics[width=1\columnwidth]{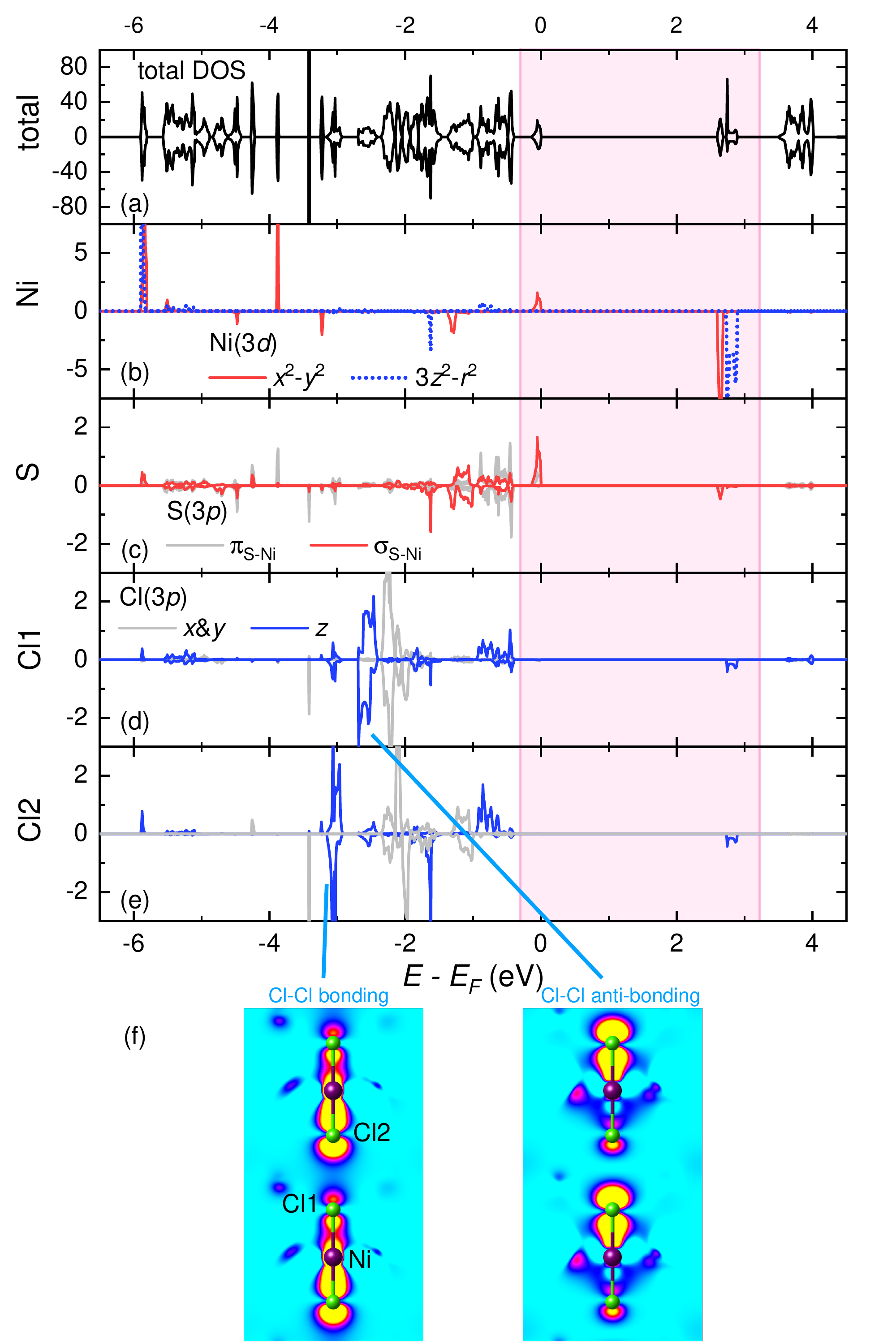}
\caption{
Density of states (DOS) of the DTN unit cell using the GGA-D3$+U$ approximation. Panel (a) is the total DOS. Panels (b)--(d) are orbital-resolved projected DOS, or PDOS. Panel (b) shows $d_{3z^{2}-r^{2}}$ (red) and $d_{x^{2}-y^{2}}$ (blue) ($e_{g}$) orbitals of Ni($3d$);  
(c) S $3p$ orbitals forming S-Ni $\sigma$ bonds and $\pi$ bonds respectively; and (d)(e) Cl($3p$) orbitals of Cl1 and Cl2 identified in (f). Positive and negative values correspond to spin-majority and spin-minority channel, respectively. The Fermi level is set to zero. The shaded region is the inner energy window for constructing the Wannier-based tight-binding Hamiltonian. The cross section in a $1\times1\times2$ supercell in (f) shows the partial charge densities of Cl-Cl bonding and anti-bonding states labeled in (d)(e). 
}
\label{fig:dos} 
\end{figure}

Experiments suggest that strong exchange-coupling along the $c$-axis is via the Ni-Cl-Cl-Ni chain\cite{nature}. To examine the interatomic interactions, we investigated the electronic structure of the DTN unit cell with antiferromagnetic ordering.
The resulting density of states (DOS) is shown in Figure~\ref{fig:dos}.
The total DOS in panel (a) shows that DTN is insulating, with a wide band gap of about $ 2.5\eV $. 
% Ni^2+ cation
In panel (b), the Ni $d_{3z^{2}-r^{2}}$ (red) and $d_{x^{2}-y^{2}}$ (blue) orbitals, often called $e_{g}$ orbitals in an octahedral crystal field, 
are fully occupied in the spin-majority channel,
and those in the spin-minority channel are almost fully empty, 
indicating a $S=1$ spin state of the $\mathrm{Ni}^{2+}$ cation.
% Cl-Cl
In panels (d) and (e), Cl-Cl bonding and anti-bonding states are identified. 
The cross sections of the partial charge densities for these states are shown in panel (f).
In the bonding state, an overlap between the two Cl($3p$) ions appears, while there is a node between the two Cl ions in the anti-bonding state.
The splitting between them is about $ 0.4\eV $. 
% bonding energy
The Cl-Cl distance is about $ 4.0 \angstrom $, much larger than the Cl-Cl bond length in $\mathrm{Cl}_{2}$, and the energy splitting is smaller than typical $pp\sigma$ bond but is still much larger than the energy of the Van der Waals interaction. 
Therefore, it reflects strong inter-molecular interactions through the Cl-Cl chain along the $c$-axis.
We note that Cl($3p$) orbitals hybridize with the Ni($d_{3z^{2}-r^{2}}$) orbitals at about $ 0.5 \eV $ below the Fermi level in the spin-up channel and about $3\eV$ above the Fermi level in the spin-down channel [Figure~\ref{fig:dos}(b)(d)(e)].
% S-Ni coupling
In the plane of $\mathrm{NiS}_{4}$ in the $\mathrm{NiS}_{4}\mathrm{Cl}_{2}$ octahedron, S atoms along S-Ni-S and along the $a$- and $b$-axes are labeled as $\mathrm{S}_{x}$ and $\mathrm{S}_{y}$, respectively.
In this case, $\mathrm{S}_{x}$($p_{x}$) and $\mathrm{S}_{y}$($p_{y}$) form $\sigma$ bonds with Ni($d_{x^{2}-y^{2}}$) so that they are labeled as S($\sigma_{\mathrm{S-Ni}}$) in panel (c).
Just below the Fermi level, Ni($d_{x^{2}-y^{2}}$) hybridizes with S($\sigma_{\mathrm{S-Ni}}$).

To understand the hybridization and the magnetic interaction pathway, we performed a unitary transformation on Bloch states to construct Wannier orbitals and the tight-binding Hamiltonian. 
We used a $1\times1\times2$ supercell with antiferromagnetic coupling along the $c$-axis, 
and Ni($e_{g}$) orbitals on each Ni ion were used as initial projections.
In this case, a total of eight Wannier functions for each spin channel are localized exclusively on Ni ions.

The inner energy window we chose, from $-0.25$ to $3.2 \eV $, is shown in Figure~\ref{fig:dos}. 
It covers both the valence and the conduction bands and so captures features of the gap for the virtual excitation in the exchange picture.
The outer energy window covers all occupied states and the unoccupied states below $5.0 \eV $. 

\begin{figure}[h]
\includegraphics[width=0.58\textwidth]{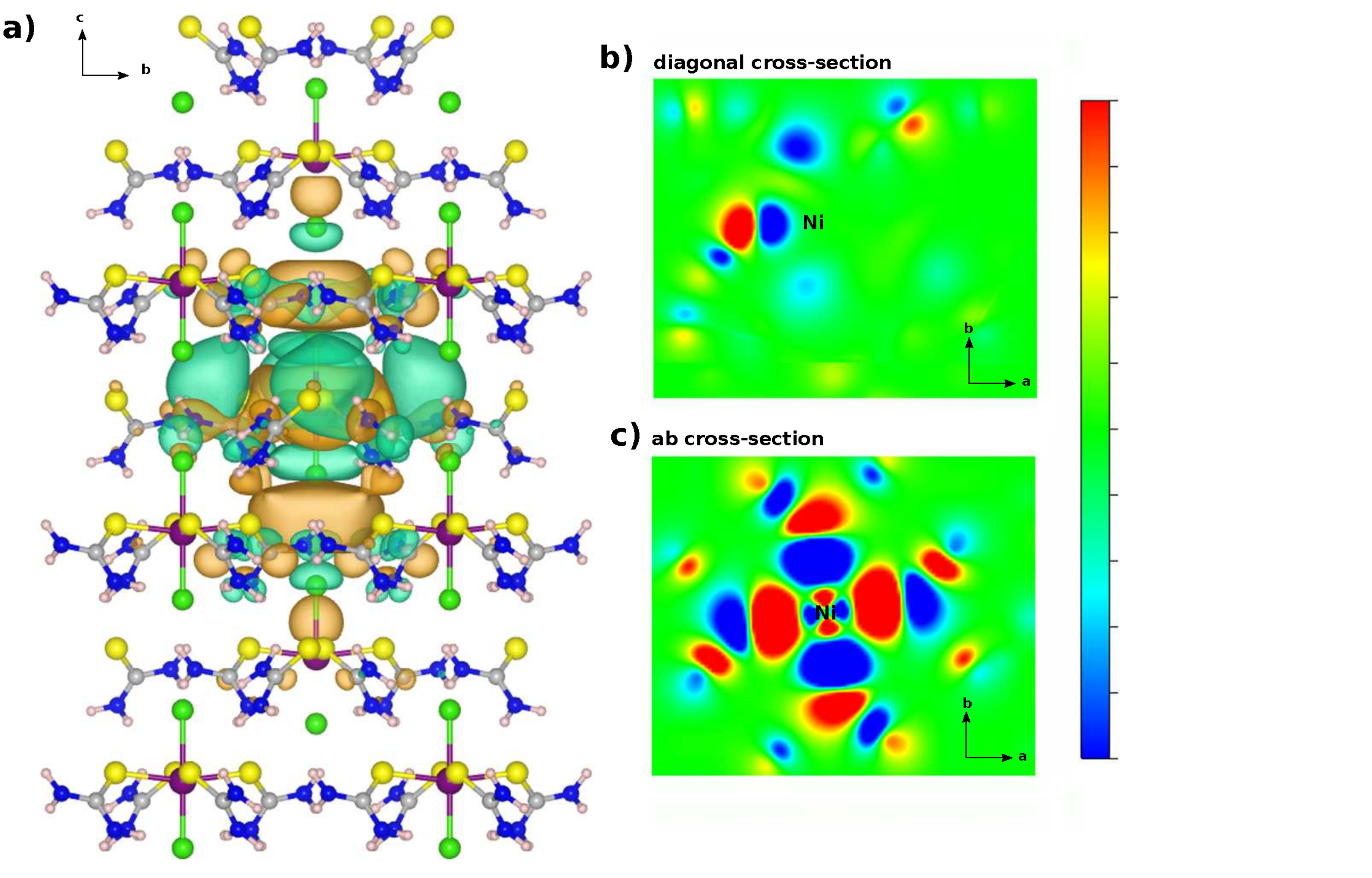}
\caption{
(a) Isosurface of a Wannier function for the AFM $c$-axis supercell, indicating hybridization of Ni and Cl orbitals leading to a superexchange path along the Ni-Cl-Cl-Ni chain; (b) Wannier orbital in a cross section containing two Ni ions along a diagonal line of the lattice; and (c) in the $ab$ plane of the lattice.
}
\label{fig:wannier} 
\end{figure}

The isosurface in Figure~\ref{fig:wannier} shows one of the Wannier orbitals, which has features of Cl($p_{z}$), S($\sigma_{\mathrm{S-Ni}}$) and Ni($e_{g}$). 
These characteristics signal strong hybridization.
The corresponding matrix elements, in units of eV, including nearest-neighbor hopping $t$ and the on-site energy $\varepsilon$ in the WF basis, are 

\begin{widetext}
\begin{equation}
\begin{array}{c|rrrrrrrr}
\mathcal{H}_{mn}~(\mathrm{eV})~ & \left|1_{3z^{2}-r^{2}}^{o}\right\rangle  & \left|1_{x^{2}-y^{2}}^{o}\right\rangle  & \left|2_{3z^{2}-r^{2}}^{u}\right\rangle  & \left|2_{x^{2}-y^{2}}^{u}\right\rangle  & \left|3_{3z^{2}-r^{2}}^{o}\right\rangle  & \left|3_{x^{2}-y^{2}}^{o}\right\rangle  & \left|4_{3z^{2}-r^{2}}^{u}\right\rangle  & \left|4_{x^{2}-y^{2}}^{u}\right\rangle \\
 \hline
\left\langle 1_{3z^{2}-r^{2}}^{o}\right| & -5.540 & 0.000 & \mathbf{-0.031} & 0.000 & 0.000 & 0.023 & 0.000 & 0.000\\
\left\langle 1_{x^{2}-y^{2}}^{o}\right| &  & -0.287 & 0.000 & 0.001 & 0.002 & 0.004 & 0.001 & 0.000\\
\left\langle 2_{3z^{2}-r^{2}}^{u}\right| &  &  & 2.570 & 0.000 & 0.005 & -0.006 & 0.005 & -0.008\\
\left\langle 2_{x^{2}-y^{2}}^{u}\right| &  &  &  & 2.405 & -0.008 & -0.002 & -0.004 & 0.000\\
\left\langle 3_{3z^{2}-r^{2}}^{o}\right| &  &  &  &  & -5.547 & 0.000 & \mathbf{-0.031} & 0.000\\
\left\langle 3_{x^{2}-y^{2}}^{o}\right| &  &  &  &  &  & -0.287 & 0.000 & 0.001\\
\left\langle 4_{3z^{2}-r^{2}}^{u}\right| &  & h.c. &  &  &  &  & 2.570 & 0.000\\
\left\langle 4_{x^{2}-y^{2}}^{u}\right| &  &  &  &  &  &  &  & 2.405
\end{array}\label{eq:TB}
\end{equation}
\end{widetext}

The position labels 1 to 4 of the WFs are the same as those in Figure~\ref{fig:supercell}. 
The subscripts $3z^{2}-r^{2}$ and $x^{2}-y^{2}$ represent the $3z^{2}-r^{2}$-like and $x^{2}-y^{2}$-like WFs centered on each Ni. 
Superscripts $o$ and $u$ refer to WFs with negative and positive on-site energies, respectively.

In particular, the largest hopping $t$ between occupied and unoccupied WFs is $-0.031\eV$ (indicated in bold in Eq.~\ref{eq:TB}), which comes from two $3z^{2}-r^{2}$-like WFs of neighboring unit cells along the $c$-axis.
Considering the difference in on-site energies, $\Delta$, of associated WFs on Ni ions along the $c$-axis, we can estimate the magnitude of their exchange interaction by $J_{c} \sim 4t^{2}/{\Delta}$, to be about $ 0.474\meV $ ($5.5 \K$).
Other hopping elements between WFs centered on Ni ions with distinct occupied states are much smaller (around $-0.001\eV$ or less), indicating a smaller exchange interaction along the diagonal direction and the $a/b$-axis directions. 
Overall, our results show that, although our estimate is of the right order of magnitude using a tight-binding model in the Wannier basis, using energetics and fitting to a Heisenberg model provides better $J$ coupling values compared to experiment. 
An alternative WF-based Hamiltonian using atomic-centered projection [See Supplemental materials for details] further confirms the strong hopping between intermolecular Cl($p_{z}$)-Cl($p_{z}$) orbitals that plays a key role in the superexchange interaction.
What we gain from our Wannier analysis\cite{Ce3Mn8nature} is more insight into the orbital contributions to the superexchange pathway.

\subsection{Magnetic anisotropy} 

\begin{figure}
    \includegraphics[width=1\columnwidth]{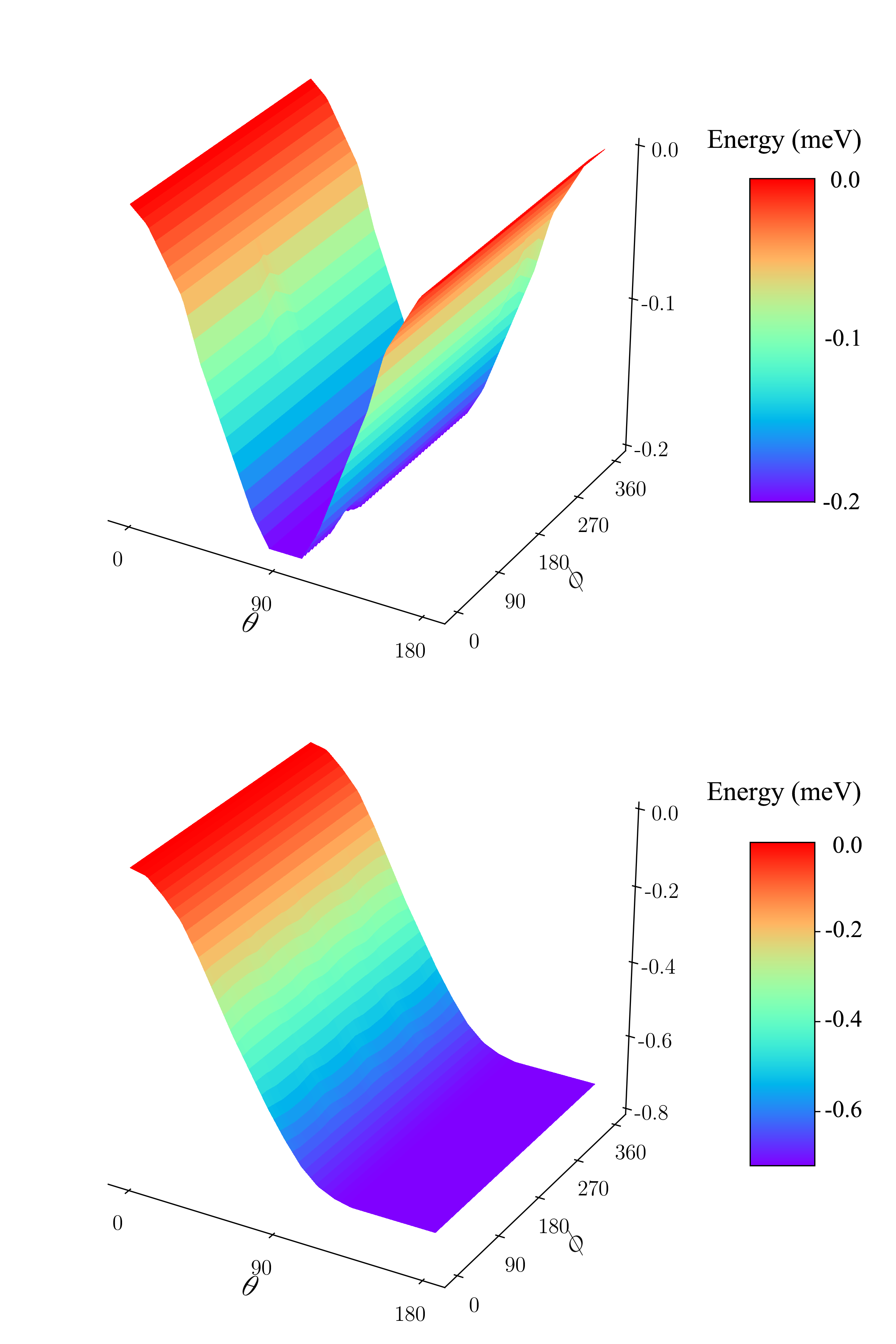}
    \caption {
    Relative total energy as a function of the angle between the $c$- axis and the spin of the Ni ion for (Top) the DTN molecule and (Bottom) the DTN unit cell with the other inequivalent Ni ion spin fixed along the $c$-axis. 
    The angles $\theta$ and $\phi$ refer to the orientation angles in spherical coordinates.
    }
    \label{fig:map}
\end{figure}

%\MY{COMMENT: I've adjusted this section a little to distinguish between the energy maps calculated initially to just see anisotropy symmetry and the actual calculation of MAE that we compared to experiment. I proofread the paper again and it seems to me the part that is repeated at the end of section III C (magnetic anisotropy) -- comparing effects of $U$ on MAE -- is more appropriate in section II C (effects of $U_{eff}$)}
Besides the exchange coupling $J$, another important interaction in Eq.~(\ref{eq:hs}) is the magnetic anisotropy. 
To study the magnetic anisotropy in DTN, we performed spin-constrained DFT calculations including the spin-orbit interaction and non-colinear magnetic moments for both an isolated DTN molecule and a DTN molecular crystal. We first look at the isolated molecule, which is placed in a $ 20 \angstrom \times 20 \angstrom \times 20 \angstrom $ supercell with adequate vacuum space, and only the $\Gamma$ point in reciprocal space is calculated. The functional used in calculations is GGA-D3. %\HPC{why are we interested in isolated molecule?}\MY{I think the main thing this showed was the easy axis. The crystal showed ground state is the AFM configuration.}
By setting a series of constrained directions of Ni spin, the total energies $E(\theta,\phi)$ as a function of direction angles are obtained.
In the crystal unit cell with two Ni atoms, the corner Ni spin is fixed to be aligned along the $c$-axis and a different spin direction for the body-centered Ni is sampled using spherical coordinates. 
The resulting relative energy maps for $E(\theta,\phi)$ are shown in  Figure~\ref{fig:map}.
In the case of a single DTN molecule, the lowest energy is at $\theta=90^{\circ}$, the in-plane direction, and the highest energies are at $\theta=0^{\circ}$ and $180^{\circ}$, when the spins are along the $c$-axis. 
Since the energy as a function of $\phi$ is almost invariant, 
this indicates that the $ab$-plane is the easy axis, with little $\phi$ dependence, and the $c$-axis is the hard axis.
The energy difference $E(0^{\circ},\phi)-E(90^{\circ},\phi)$ is $0.20\meV$ ($2.32 \K$).
This result shows a uniaxial magnetic anisotropy of the DTN molecule.
For the molecular crystal, $E(0^{\circ},\phi)$ is the highest while $E(180^{\circ},\phi)$ is the lowest. 
The energy difference $E(0^{\circ},\phi)-E(180^{\circ},\phi)$ is $0.59\meV$ ($6.84 \K$).
Since the spin of the corner Ni is fixed along the $c$-axis, the direction of the central Ni spin with lowest energy corresponds to antiferromagnetic spin-ordering with the corner Ni spin. 

The definition of $D$, the zero-field splitting, is the energy difference between states with $S_{z}=\pm1$ and $S_{z}=0$ for an $S=1$ system. DFT calculations can only give the quantum state $S_{x}=\pm1$, instead of $S_{z}=0$, when the spin is constrained along $a$-axis. 
To calculate the magnetic anisotropy for the unit cell with two Ni atoms, we take the energy difference between configurations where the spins in one configuration are aligned along the $c$-axis and along the $a$-axis for the other configuration. 
The energy difference $E(0^{\circ},\phi)-E(90^{\circ},\phi)$ per Ni by DFT is just half the magnitude of $D$.
Therefore, $D=2[E(0^{\circ},\phi)-E(90^{\circ},\phi)]$ for DTN.
The experimental value. is $D = 0.774 \meV$ ($8.98 \K$).

For the DTN crystal, we performed both GGA-D3 and GGA-D3+$U$ with $U_{\eff}=6.0$~\eV.
Using the GGA-D3 functional, we underestimate $D$ to be $ 0.54 \meV$  ($6.26 \K$).
A GGA-D3+$U$ calculation gives $D = 0.78 \meV$ ($9.10 \K$), consistent with the experimental results. 
We also examined $U =7.0 \eV$ and $U-J$ with $U=4.9 \eV$ and $J= 0.52 \eV$ calculated by cRPA and found that $D$ is insensitive corresponding to a $D$ value of $0.78 \meV$ for both.
%The angular dependence of the energy curves in Figure~\ref{fig:map} remains unchanged. 
It is worth noting that for heavy elements that involve $f$ electrons, a nonspherical $U-J$ scheme may be needed \cite{RN3270}. 
% \MY{to Jie-Xiang: I've read over everything above this point and made small adjustments to wording where I thought it need. Look over it if you can and let me know how it sounds. There's a new ME effect section below and I'm working on trying to collect data for tables and plots.}
% \MY{Since using the dispersion corrected GGA-D3 XC function allows us to accurately} predict DTN's structure, we went on to calculate basic magnetic properties of the molecular crystal \MY{which contains two $S=1$ $Ni^_{2+}$ ions in its unit cell}. 
% We first performed a series of spin-constrained DFT calculations on the DTN molecular unit itself, allowing the total energy, $E(\theta,\phi)$, to be a function of spherical coordinates and producing a energy map  (Figure~\ref{fig:map}). 
% From the top figure we see \MY{there is an easy-plane at} $\theta=90^{\circ}$, with little $\phi$ dependence, and the uniaxial symmetry of the molecule. 
% The bottom figure is the energy map produced using the DTN unit cell that contains two in-equivalent Ni ions. 
% In this unit cell we have a Ni ion at the corner of our box and one body-centered Ni. 
% To produce this energy map we fixed the corner Ni's spin to lie along the $c$-axis of the cell and sampled different spherical coordinates for the central Ni. 
% This shows the ground state is anti-ferromagnetic (AFM), though energy differences between neighboring configurations are small, indicating the paramagnetic nature of DTN at low temperatures and under zero field. 

%\subsection{Magnetic anisotropy}
% In the results that follow we make comparisons between two different approximations, namely, DFT calculations performed using GGA and GGA+$U$. Experimentally it is shown that the zero-field splitting, $D$, between $S_z=0,\pm1$ is 8.98 K. 

A fuller investigation of magnetic anisotropy, including the effects of substitutions of Cl ions and ligands, will be left for future work. 
Earlier results suggest a decrease in magnetic anisotropy as the substitute ion gets larger\cite{jie's-paper}.

%We also studied the effects of different %substitutions of the Cl ion on the magnetic %anisotropy. What we see a decrease in mae as the %substitute ion gets larger, i.e. from Cl to Br to %I. Further First-Principles studies of doping %effects will be left for future work. 
%\HPC{I thought that study is in the paper with XGZ's group}\MY{You're right, I could reference or just remove it? This is an old paragraph that seemed a little out-of-place.}
% mae with doping table
% \begin{table}[h]
%     \centering
% \begin{tabular}{ |P{2cm}|P{2cm}|}
%   \hline
% System  & MAE (meV) \\ 
%  \hline
%  DTN     & 0.54   \\
%  Br-DTN  & 0.39   \\
%  I-DTN   & 0.22   \\
%  \hline
% \end{tabular}
% \caption{Table 1. Magnetic anisotropy for pure DTN and substituted DTN. One Cl ion is % replaced for each structure.}
% \end{table}

\subsection{Magneto-electric effect}
%ME effect

\begin{figure}
\includegraphics[width=1\columnwidth]{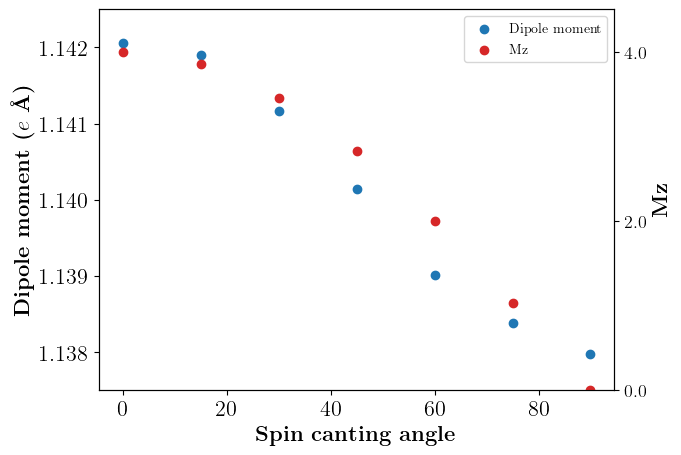}
\caption{
Magneto-electric coupling in DTN. The electric dipole moment (blue) and the $z$-component of the magnetization (red) are plotted as a function of the spin canting angle of the spin moment on the Ni cation.}
\label{fig:me}
\end{figure}

With an understanding of the magnetic interactions in DTN we turn to the magneto-electric (ME) effect. To investigate this effect we again performed spin-constrained DFT calculations, including the spin-orbit interaction and non-colinear magnetic moments. This was done for the DTN molecular crystal with two Ni atoms and using the GGA-D3 functional and its corresponding relaxed structure. Constraining the moment direction of each Ni relative to the $c$-axis allowed us to simulate what magnetic configurations would result when a magnetic field is applied parallel to the $c$-axis of the molecular crystal. The spin canting angle in Figure~\ref{fig:me} is equivalent to the polar angle relative to the $c$-axis. A spin-canting angle of $0^{\circ}$ corresponds to spins that are parallel to the $c$-axis and representing the spin-polarized state at high magnetic fields. Similarly, a $90^{\circ}$ angle corresponds to spins lying in the $ab$-plane and the $XY$-AFM state. For each configuration we calculated the electric dipole moment of the crystal within the modern theory of polarization framework\cite{polarization}. What we see in Figure~\ref{fig:me} is that as the canting angle $\theta$ increases and the spins approach the $ab$-plane, there is a correlated decrease in the electric dipole moment (blue) of the crystal together with the expected decrease in $M_z$ (red). This indicates the presence of an ME coupling within the DTN molecular crystal.

\begin{figure}
\includegraphics[width=1\columnwidth]{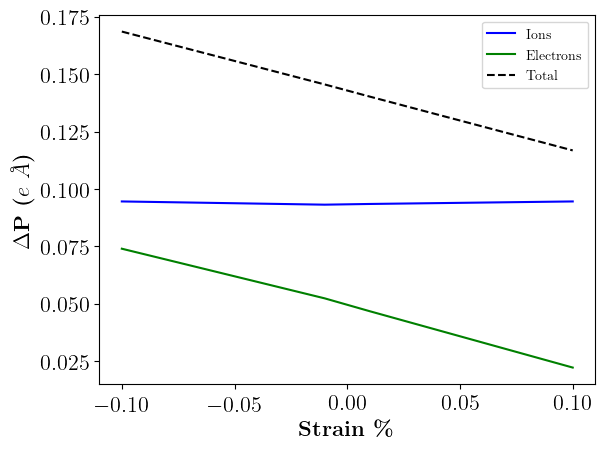}
\caption{ 
Ionic and electronic contributions to the change in polarization of the DTN unit cell under applied strain.}
\label{fig:me-strain}
\end{figure}

%ME strain
To probe the mechanism responsible for the ME coupling we investigate what effect straining the molecular crystal has on the electronic polarization. The literature suggests magnetic field-induced strain, \textit{i.e.} magnetostriction, changes the unit cell and a reorientation of the electrically polar ligands may be responsible for the change in polarization\cite{magnetostriction}. To test this, we used the DTN molecular crystal and the GGA-D3 functional structures from our ME calculations with colinear spins in an AFM configuration and no spin-orbit interaction. We applied a tensile strain on the DTN unit cell along the $c$-axis of up to 0.1\% while correspondingly compressing the $a$, $b$ lattice constants, keeping the volume constant, and allowed the ions to relax to minimize the total energy. Similarly, we applied a compression to the $c$ lattice constant while expanding the $a$, $b$ constants. We calculated the polarization as a function of applied strain. We distinguish the change in polarization relative to the unstrained system, $\Delta$P, resulting from ionic motion (blue) from that due to changes in electronic density (green) in Figure~\ref{fig:me-strain}. The results indicate that the main contribution to the change in polarization comes from subtle changes in electronic density rather than a reorientation of the thiourea ligands. An analysis of the molecular crystal structure shows little variation in the Cl-Ni-S bond angles or bond lengths, supporting this result.
%\HPC{Somewhere, define $ \Delta P = $ (something).  The units on the vertical axis of the figure should be italic $e$, roman {\AA}, $ e\angstrom $, the opposite of what is there now}

\begin{figure}
\includegraphics[width=1\columnwidth]{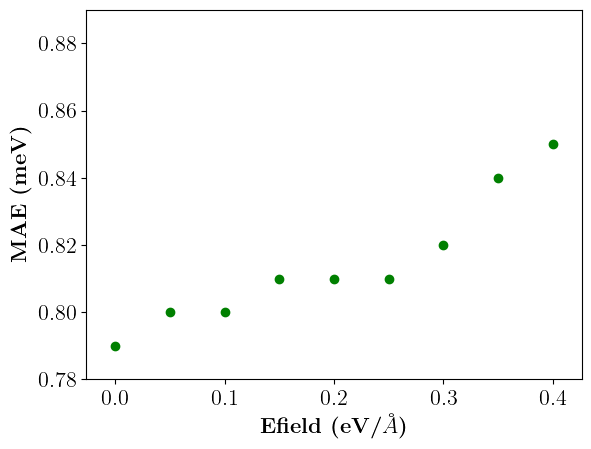}
\caption{
The dependence of the magnetic anisotropy of bulk DTN on an electric field applied parallel to the $ab$-plane.}
\label{fig:efield}
\end{figure}

To conclude our study of the ME effect in DTN, we investigate what effect applying an electric field has on the magnetic properties of the molecular crystal. Specifically, we investigate the magnetic anisotropy (MAE) as a function of electric field strength where the electric field is aligned parallel to the $a$-axis and perpendicular to the 1D-chain. In these calculations we used the DTN molecular crystal unit cell structure with two Ni atoms and the GGA-D3+$U$ exchange-correlation functional with $U_{\eff}=6 \eV$. As with our previous MAE calculations, we performed noncolinear magnetic DFT calculations including spin-orbit interactions and obtained the energy difference between two configurations of spins aligned $90^{\circ}$ relative to one another. The results show that by increasing the strength of the electric field, we are able to increase the MAE of bulk DTN (Figure~\ref{fig:efield}). This indicates the presence of a reverse coupling in this molecular crystal, not observed experimentally, showing that the magnetic properties may be tuned by electronic means.

%Further investigations (Figure~\ref{fig:me-strain}) shows little change in polarization due to changes in the ionic structure and that the main contribution comes from subtle changes in electronic degrees of freedom. This result is an indication that re-orientation of the ligands in fact does not happen.

%\begin{figure}[lh]
%\centering
%\includegraphics[width=0.42\textwidth]{me-effect-2.png}
%\caption{\label{Figure} 
%Linear relationship between the electric dipole moment and $M_z^2$}
%\end{figure}

\section{Conclusion}

Our aim in this paper has been an attempt to model DTN within a DFT framework. 
This approach not only provides a way to gain insight into the mechanisms responsible for structural, electronic and magnetic properties, but also allows us to make accurate predictions while minimizing the computational cost. 
We were able to show that the inclusion of the GGA-D3 correction substantially improves the structure of the DTN molecular crystal, providing accurate lattice constants, bond lengths and bond angles.
Adding the Hubbard term to the standard GGA functional allowed us to improve our modelling of the magnetic interactions in DTN, accurately predicting the magnetic anisotropy and exchange coupling constants and providing an additional $J_\diag$ coupling. 
The quasi-1D nature of DTN was indicated by the hybridization of Ni($d_{3z^{2}-r^{2}}$) and Cl-Cl anti-bonding orbitals via a virtual superexchange path along the $c$-axis. 
Finally, we showed the presence of a ME coupling and predict an increase in magnetic anisotropy with the application of an electric field.
The results of this study are a first step towards understanding the detailed mechanisms involved in the magneto-electric coupling in DTN. 
Its presence in functional organic quantum magnets suggests the possibility of future applications where the magnetic properties of materials may be fine-tuned through more feasible structural and electronic means. 
Yet to be explored in detail is the nature of Bose-Einstein condensates (BEC) in organic quantum magnets\cite{nature} and the important role the ME effect plays in this state. 
xWe hope this work puts forward the idea that these fundamental properties may be investigated from first principles.

~

\textbf{Acknowledgments:} This work was supported as part of the Center for Molecular Magnetic Quantum Materials, an Energy
Frontier Research Center funded by the U.S. Department
of Energy, Office of Science, Basic Energy Sciences
under Award No.~DE-SC0019330. Computations were performed
at NERSC and UFRC.

\appendix
\renewcommand{\thefigure}{\thesection\arabic{figure}}
\renewcommand{\thetable}{\thesection\arabic{table}}

\bibliography{main}
\end{document}